\documentclass[11pt]{article}
\usepackage{moriond,epsfig}

\def\hmpc{\rm \,h^{-1}\,Mpc}

\def\xir{$\xi(r)$\ }

\def\wp{$w_p(r_p)$\ }

\def\n_med{{\left<n\right>}}

\def\begc{\begin{center} }
\def\endc{\end{center} }
\def\begf{\begin{figure} }
\def\endf{\end{figure} }

\def\j3{{J_3}}

\newcommand{\mincir}{\raise -2.truept\hbox{\rlap{\hbox{$\sim$}}\raise5.truept
\hbox{$<$}\ }}
\newcommand{\magcir}{\raise -2.truept\hbox{\rlap{\hbox{$\sim$}}\raise5.truept
\hbox{$>$}\ }}
\newcommand{\siml}{\raise -2.truept\hbox{\rlap{\hbox{$\sim$}}\raise5.truept
\hbox{$<$}\ }}
\newcommand{\simg}{\raise -2.truept\hbox{\rlap{\hbox{$\sim$}}\raise5.truept
\hbox{$>$}\ }}
\newcommand{\be}{\begin{equation}}
\newcommand{\ee}{\end{equation}}
\newcommand{\ba}{\begin{eqnarray}}
\newcommand{\ea}{\end{eqnarray}}
\newcommand{\brr}{\begin{array}}
\newcommand{\err}{\end{array}}
\newcommand{\bc}{\begin{center}}
\newcommand{\ec}{\end{center}}

\def\be{\begin{equation}}
\def\ee{\end{equation}}
\def\bea{\begin{eqnarray}}
\def\eea{\end{eqnarray}}

\begin{document}
\vspace*{4cm} 
\title{STUDYING THE EVOLUTION OF LARGE-SCALE STRUCTURE
  WITH THE VIMOS-VLT DEEP SURVEY}

\author{ LUIGI GUZZO$^{1,2}$ and the VVDS Consortium$^*$ }

\address{$^1$INAF - Osservatorio Astronomico di Brera, 
Milan, 
Italy}

\maketitle\abstracts{ The VIMOS-VLT Deep Survey (VVDS) currently
  offers a unique combination of depth, angular size and number of
  measured galaxies among surveys of the distant Universe: $\sim
  11,000$ spectra over $0.5$ deg$^2$ to $I_{AB}=24$ ({\it VVDS-Deep}),
  $35,000$ spectra over $\sim 7$ deg$^2$ to $I_{AB}=22.5$ ({\it
    VVDS-Wide}).  The current ``First Epoch'' data from VVDS-Deep
  already allow investigations of galaxy clustering and its dependence
  on galaxy properties to be extended to redshifts $\sim 1.2-1.5$, in
  addition to measuring accurately evolution in the properties of
  galaxies up to $z\sim 4$.  This paper concentrates on the main
  results obtained so far on galaxy clustering.  $L_B^*$ galaxies at
  $z\simeq 1.5$ show a correlation length $r_0=3.6\pm 0.7$.  As a
  consequence, the linear galaxy bias at fixed luminosity rises over
  the same range from the value $b_L\simeq 1$ measured locally, to
  $b_L=1.5 \pm 0.1$.  The interplay of galaxy and structure evolution
  in producing this observation is discussed in some detail.  Galaxy
  clustering is found to depend on galaxy luminosity also at $z\simeq
  1$, but luminous galaxies at this redshift show a significantly
  steeper small-scale correlation function than their $z=0$
  counterparts.  Finally, red galaxies remain more clustered than blue
  galaxies out to similar redshifts, with a nearly constant relative
  bias among the two classes, $b_{rel}\simeq §1.4$, despite the rather
  dramatic evolution in the color-density relation over the same
  redshift range.  }  

\renewcommand{\thefootnote}{\fnsymbol{footnote}}
\footnotetext{$^2$\sl Visiting Scientist, 
European Southern Observatory \& Max-Planck
  Gesellschaft, Garching, D}
\footnotetext[1]{{\sl The VVDS Consortium:} 
\tiny O. Le F\`evre (LAM Marseille),
G. Vettolani (INAF-IRA Bologna),
C. Adami (LAM Marseille),
S. Arnouts (LAM Marseille),
S. Bardelli  (INAF-OA Bologna), 
M. Bolzonella  (INAF-OA Bologna), 
M. Bondi (INAF-IRA Bologna),
A. Bongiorno (Univ. Bologna),
D. Bottini (INAF-IASF Milano),
J. Brinchmann (Porto),
A. Cappi    (INAF-OA Bologna),
S. Charlot  (IAP Paris),
P. Ciliegi    (INAF-OA Bologna),  
T. Contini (Toulouse),
O. Cucciati (INAF-OA Brera),
S. De la Torre (LAM Marseille),
S. Foucaud (Nottingham),
P. Franzetti (INAF-IASF Milano),
B. Garilli (INAF-IASF Milano),
I. Gavignaud (AIP Potsdam),
L. Guzzo (INAF-OA Brera),
O. Ilbert (IfA Hawaii),
A. Iovino (INAF-OA Brera),
F. Lamareille (INAF-OA Bologna), 
V. Le Brun (LAM Marseille),
D. Maccagni (INAF-IASF Milano),
H.J. McCracken (IAP Paris),
B. Marano  (Univ. Bologna),
C. Marinoni (Univ. Marseille),
A. Mazure (LAM Marseille),
Y. Mellier (IAP Paris),
B. Meneux (INAF Milano),
R. Merighi   (INAF-OA Bologna), 
P. Merluzzi (INAF-OA Napoli),
S. Paltani (ISDC Geneva),
R. Pell\`o (Toulouse),
J.P. Picat (Toulouse),
A. Pollo (LAM Marseille \& Cracow),
L. Pozzetti (INAF-OA Bologna), 
M. Radovich (INAF-OA Napoli),
R. Scaramella (INAF-OA Roma),
M. Scodeggio (INAF-IASF Milano),
S. Temporin (INAF-OA Brera),
L. Tresse (LAM Marseille),
D. Vergani (INAF-IASF Milano),
C.J. Walcher (LAM Marseille)
G. Zamorani (INAF-OA Bologna), 
A. Zanichelli (INAF-IRA Bologna),
E. Zucca    (INAF-OA Bologna),
}

\section{Introduction}

During the last few years, our knowledge of the large-scale structure
of the Universe has reached an exquisite level of accuracy.  This has been
the result of extensive surveys of galaxy redshifts, as the
2dFGRS (Colless et al. 2001)
and SDSS (Adelman-McCarthy et al. 2006),
that constructed
complete flux-limited samples of several hundred thousand galaxies in the
local ($z<0.2$) Universe, allowing detailed statistical analyses to be
performed (e.g. Cole et al. 2005, Eisenstein et al. 2005).

The extension of similar systematic studies of large-scale structure
over significant volumes at redshifts approaching unity, is on
the other hand relatively recent.  Building upon the
pioneeristic efforts over very small areas
(Broadhurst et al. 1988, Le Fevre et al. 1996, Yee et al. 2000),
large ($N > 10,000$)
systematic surveys of galaxy redshifts reaching faint flux limits
($AB$ magnitudes fainter than $\sim 22.5$) and covering {\it
at the same time} areas of the order of or larger than 1 square
degree, have become 
possible only in very recent years, thanks to the construction of
highly-multiplexed spectrographs coupled to 8-meter class
telescopes.  There are currently three such projects ongoing, the
VIMOS-VLT Deep Survey, whose clustering results will be the subject of
this paper, the DEEP2 survey at Keck (Coil et al. 2006)
and the zCOSMOS survey, again using VIMOS at the VLT (Lilly et al. 2007).



%
\begin{figure}
\begin{center}
\epsfxsize=15cm  
\epsfbox{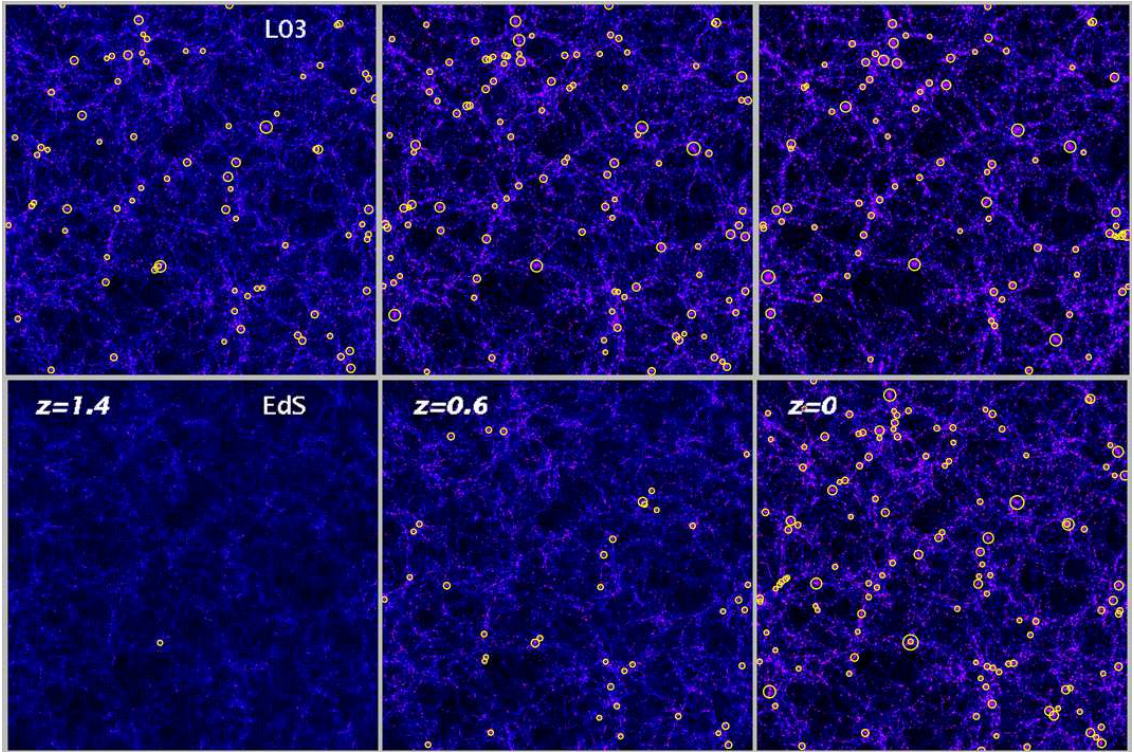}  
\caption{The growth of structure in the Universe simulated through two
  N-body experiments, both realistic and normalized as to reproduce
  the observed clustering at the current epoch: a flat low-density
  model (top, $\Omega_m=0.3$, $\Omega_\Lambda=0.7$) and an
  Einstein--de-Sitter model (bottom, $\Omega_m=1$).  Each of the three
  redshift snapshots shows a comoving slice of 75 h$^{-1}$ Mpc
  thicknes and 250 h$^{-1}$ Mpc side.  Superimposed on the gray-scale dark
  matter distribution, circles mark galaxy clusters
  with masses corresponding to an $X$--ray temperature $T_X>3$ keV, with
  size proportional to $T_X$ (figure from Borgani \& Guzzo 2001).  
}
\label{Borgani-Guzzo}
\end{center}
\end{figure}
%
Figure~\ref{Borgani-Guzzo} shows explicitly which was the main
original motivation of astronomers for trying and measure the
evolution of galaxy clustering: the way structure grows is a sensitive
function of the cosmological model (from Borgani \& Guzzo 2001).
The top and
bottom rows in the figure show snapshots at three different redshifts
from two N-body simulations run with different cosmological initial
conditions, as described in the caption. Both are normalized as to
reproduce the observed clustering at the current epoch.
The yellow circles mark the position and mass of galaxy clusters,
identified in the simulation as mass concentrations that would shine
in X-rays with an equivalent temperature $T_X>3$ keV.  The evolution
of large-scale structure with redshift is dramatically different in
the two simulations: in the Einstein-DeSitter model at the bottom
(where the flat geometry is provided entirely by matter,
$\Omega_M=1$), a shortage of massive clusters is already observed at
redshifts as low as 0.6.  On the contrary, the low-$\Omega_M$ model
shows almost no evolution.  The results from deep surveys of X-ray
clusters  are consistent with
this latter picture, showing only a mild decline of the bright end 
of the cluster X-ray luminosity function for $z>0.7$ (see 
Rosati et al. 2002 
for a review).  
Using the evolution of structure sampled via the abundance of clusters at
different epochs, therefore, one can obtain fairly robust estimates of
the mean density of matter\footnote[2]{More
  precisely, the evolution of the cluster mean density depends both on
  $\Omega_M$ and on the normalization of the power spectrum of
  fluctuations, traditionally espressed as the {\it rms} fluctuation
  in $8 \hmpc$ spheres, $\sigma_8$. },  $\Omega_M\simeq 0.3$ (Borgani
et al. 2001).

These results from X-ray cluster are made possible by one specific
feature: at this level, clusters can be considered as rather simple
dark matter halos, for which we have well-calibrated relationships
relating the halo total mass and the observed X-ray luminosity.  Given
a cosmological model and thus a power spectrum of fluctuations, we can
robustly predict the number density of DM halos above a given mass at
any epoch, via the Press-Schechter (1974) theory
and subsequent
refinements (e.g. Sheth \& Tormen 1999),
or via analytic fits to
N-body simulations like those of Fig.~\ref{Borgani-Guzzo}
(e.g. Jenkins et al. 2001).
This can then be translated into the
expected number of clusters above a given X-ray luminosity, and
compared to observations.

The situation with using galaxies as tracers of structure evolution is
significantly more complicated: understanding the relationship between
dark matter halos and the baryonic component we actually detect in the
form of galaxies is in fact one of the major challenges of modern
cosmology.  Theoretical efforts, in view of the lack of a complete
physical descriptions of the large number of nonlinear processes
leading to galaxy formation within a dark-matter halo, have
necessarily to rely upon a number of well-motivated physical recipes.
These are used to describe statistically phenomena like cooling, star
formation and heating in the baryonic component, as a function of the
properties of the parent halo and, possibly, the surrounding
environment.  Such semi-analytic scheme is usually applied to the
history and distribution of halos (the ``merging tree''), obtained via
a Montecarlo Press-Schechter approach or from n-body simulations
(e.g. Somerville \& Primack 1999, Kauffmann et al. 1999, Lacey \&
Cole 1993). 
Thus, the
effect of the fundamental cosmological parameters that shows up so
clearly in the example of Fig.~\ref{Borgani-Guzzo}, 
when we come to galaxies is in principle shielded by the
myriad of astrophysical processes related to gas and stars.

Nevertheless, the motivation for extensive deep galaxy redshift
surveys is not diminished: one wants not only to measure the
clustering properties as a function of redshift, but also how these
depend, at different cosmic epochs, on physical properties (as
e.g. luminosity, star-formation rate, color, morphological type) that
are themselves evolving with redshift.  The goal becomes thus also
understanding how these may be connected, more or less directly, to
the mass of the parent halo and/or to the surrounding environment.
Changing our initial perspective, therefore, we may say that the main
motivation for deep galaxy surveys is not to try and measure
cosmological parameters via the growth of DM structure (which is
difficult to disentangle from the aforementioned astrophysical
effects), but rather learn how galaxies have been forming within DM
halos. The hope is that at the end of the day we can close the circle
and better understand how to use them to do cosmology\footnote[3]{Throughout
  the paper we shall use $\Omega_m = 0.3$ and $\Omega_{\Lambda} =
  0.7$, with the Hubble constant usually parameterised via h$=H_0/100$
  to ease comparison with previous works; a value $H_0 = 70$ km
  s$^{-1}$ Mpc$^{-1}$ is used when computing absolute magnitudes.}.

\section{The VIMOS-VLT Deep Survey}

%
%
The VIMOS-VLT Deep Survey (VVDS) was designed
specifically to probe the combined evolution of galaxies and large
scale structure to $z \sim 2$ (reaching up to $z\sim 4.5$ with the
most extreme objects), by measuring $\sim 100,000$ faint galaxy
redshifts. The VVDS is built around the VIMOS multi-object
spectrograph at the ESO VLT, capable of simultaneously
collecting between 380 and 600 spectra (Le F\`evre et al. 2005).
Its source catalogue
is selected in the $I_{AB}$ band and the survey is composed by two
distinct parts with complementary science goals: {\it VVDS-Deep},
covering $\sim 1$ sq. deg. to $I_{AB}=24$ (5-hour exposures); and {\it
  VVDS-Wide}, covering more than $10$ sq. deg. to $I_{AB}=22.5$
(1-hour exposures).  
%
Virtually all results presented here are based on the ``First Epoch''
set of 6530 reliable ($>80\%$ confidence) redshifts, covering $0.49$
square degrees in the ``2-hrs field'' (F02) of VVDS-Deep.  These were
collected during the guaranteed-time observations awarded to the
Consortium for the construction of VIMOS.  The F02 field has also been
the subject of extended multi-band observations, including UBVRIK, IR
(Spitzer-Swire), radio (VLA) and X-ray (XMM). Another $\sim 33,000$
spectra have already been secured by the $Wide$ survey -- which is
still under completion -- and their analysis is undergoing.  Details
about observations, data reduction, redshift measurement and quality
assessment can be found in Le F\`evre et al. (2004; 2005).

\section{The evolution of galaxy clustering since $z\sim 1.5$}

\begin{figure}
\begin{center}
\epsfysize=8.cm\epsfbox{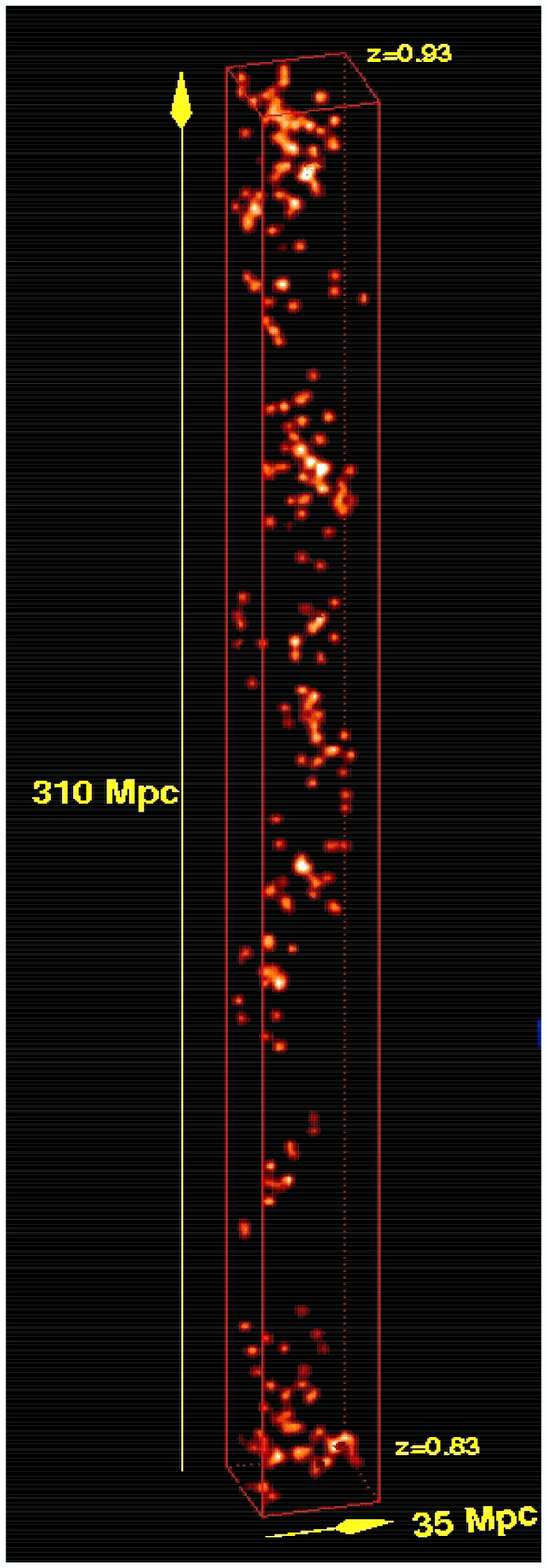}  
\epsfysize=8.cm\epsfbox{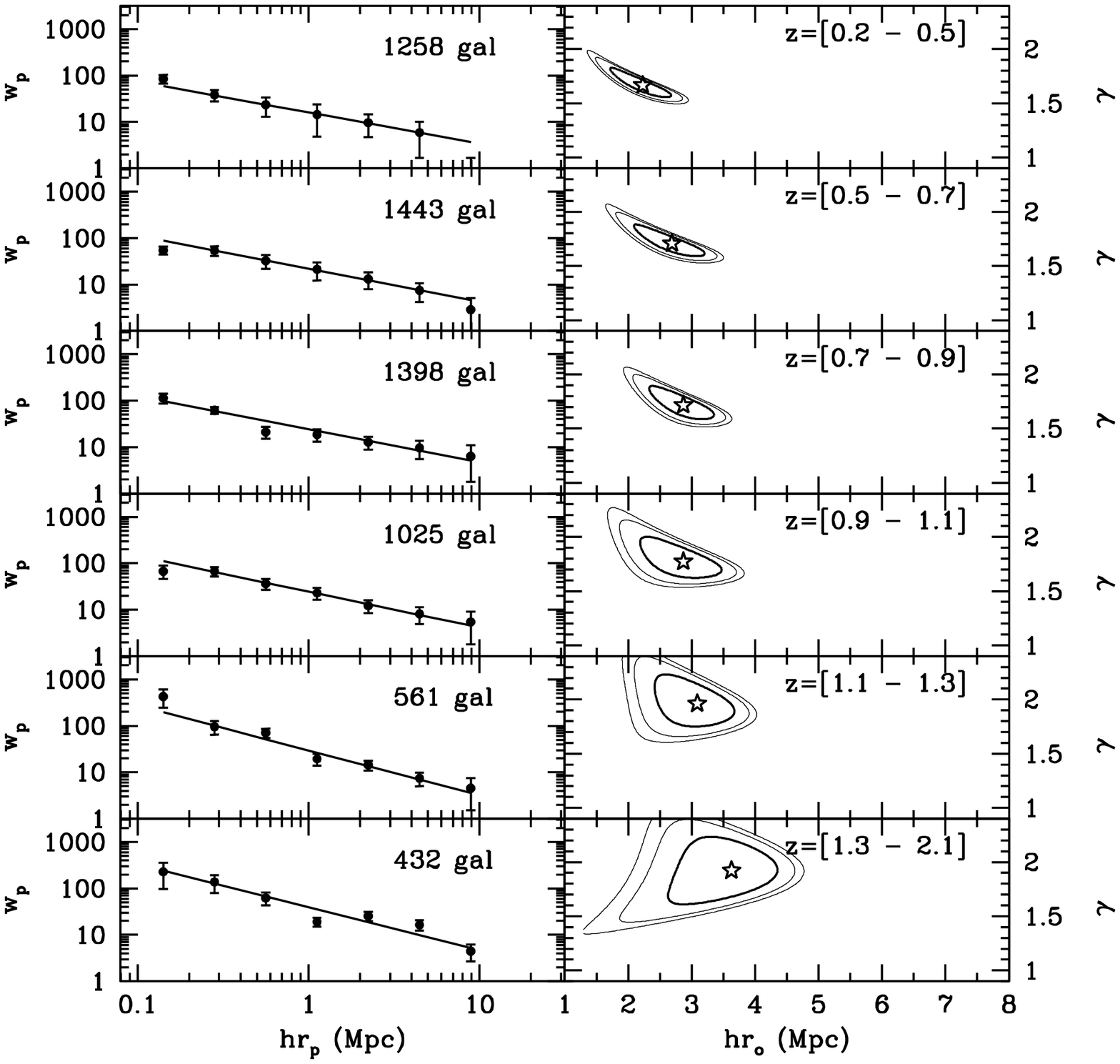}
\caption{{\it Left:} Large-scale structure at $z\sim 1$, as seen in a
small section of the VVDS light cone.  This figure (see Marinoni,
this volume) shows the true density field, smoothed on a scale of $2
\hmpc$ and corrected for the survey selection function.  {\it Right:}
Measurements of the projected function \wp in redshift slices drawn
from the full flux-limited VVDS survey, with power-law fits and the
corresponding $r_0$,$\gamma$ parameters of the real-space correlation
function. }
\label{xi_vvds_1} 
\end{center}
\end{figure}

As with any large redshift survey, the first result from the currently
available VVDS data is a cartographic map of the galaxy distribution.
The novelty is that the VVDS map is able to cover redshifts never
sampled before with similar three-dimensional accuracy.  Previous deep
surveys were limited to showing peaks and valleys in their 1D redshift
histograms, given their small area.  With the VVDS, we can for the
first time contemplate the appearance of large-scale structure at an
epoch when the Universe was about half its current age.  An example
from a cut-out section of the VVDS ``light-cone'' between $z=0.83$ and
0.93 is shown in the left panel of Fig.~\ref{xi_vvds_1}.  At this
redshift, the survey samples transverse separations of the order of
$30 \hmpc$.  Supercluster structures and ``voids'' similar to those
observed by local surveys are clearly emerging.  The reader may find
this picture somewhat familiar, being used to the similarly beautiful
light-cones derived from large numerical simulations.  However, while
looking at this picture it is worth keeping in mind that this is
indeed the real Universe and consider how these maps represent indeed
one further step forward in the ever-lasting quest by mankind to
explore and chart the surrounding world. 

In the following sections, the inhomogeneities that are evident in
the galaxy distribution will be quantified in terms of their
autocorrelation properties and compared to what we measure at $z\sim
0$.

\subsection{Real- and redshift-space correlations at $z\sim 1$}

The simplest statistic for studying clustering in the galaxy
distribution is the two-point correlation function, \xir.  This
measures the excess probability over random of finding a pair of
galaxies with separation $r$.  Since in redshift space galaxy
distances include the contribution from peculiar velocities, it is
useful to split the separation $r$ into the two components $r_p$ in
the plane of the sky, and $\pi$ along the line-of-sight.  The latter
variable, $\pi$ will thus be in this formulation the only one affected
by the distortions produced by peculiar motions.  These distortions
contain important information on cosmological parameters, 
specifically on the value of $\beta\simeq
\Omega_M^{0.6}/b$, where $\Omega_M$ is the matter density parameter
and $b$ is the so-called {\it bias} for the specific class of objects
being used (Kaiser 1987),
on which we shall not discuss here.

%
$\xi(r_p,\pi)$ is estimated by comparing the observed counts of galaxy
pairs at a given separation, with those measured from a random
distribution to which we have applied both on the sky and in redshift
the same selection function as the observed galaxies.  Additionally,
pairs are weighted as to take into account systematics effects
introduced by the instrumental set-up and observing strategy.  In the
application to the VVDS data, this procedure has been calibrated and
extensively tested using 50 VVDS-Deep mock samples, built from the
GalICS simulations (Blaizot et al. 2005).
The results have been shown
to be robust against the uncertainties in the weighting procedure
(Pollo et al. 2005).

The galaxy real-space correlation \xir can thus be recovered, free of
distortions, by projecting $\xi(r_p,\pi)$ along the line of sight
direction, $\pi$ (Davis \& Peebles 1983)
%
\be
w_p(r_p) \equiv 2 \int_0^\infty \xi(r_p,\pi) d\pi\,.
\label{wpdef}
\ee

For a power-law spatial correlation function $\xi(r) =
(r/r_0)^{-\gamma}$, the integral can be computed analytically, and the
projected correlation function can be expressed as
\begin{equation}
w_p(r_p) = r_p \left(\frac{r_0}{r_p}\right)^\gamma
\frac{\Gamma\left(\frac{1}{2}\right)\Gamma\left(\frac{\gamma-1}{2}\right)}
{\Gamma\left(\frac{\gamma}{2}\right)} \propto r_p^{-\gamma+1} \,\,\, ,
\label{wpmodel}
\end{equation}
where $\Gamma$ is Euler's Gamma function.  This model can thus be fit
to the observed $w_p(r_p)$ at different redshifts, to estimate the
values of $r_0$ and $\gamma$ that best describe the global amplitude
and slope of the underlying $\xi(r)$.  In this way, all the
information on $\xi(r)$ at a given redshift is compressed into two
numbers.  This is useful for a first compact comparison of clustering
at different redshifts.  However, important information 
is contained in the detailed shape of $w_p(r_p)$ and $\xi(r)$
(e.g. Guzzo et al. 1991, Zehavi et al. 2004).
It is therefore of interest, if the data are sufficient,
to go beyond the simple power-law approximation and compare the whole
shape of $w_p(r_p)$ at different redshifts (see \S~\ref{xi_lum}).

\subsection{Observed clustering evolution from the VVDS-Deep
  flux-limited sample}
\label{xi_1}

The right panel of Fig.~\ref{xi_vvds_1} shows the estimate of
$w_p(r_p)$ from the pure flux-limited $I_{AB}<24$ VVDS-Deep 
(Le Fevre et al. 2005), 
within bins at different redshifts, together with the
corresponding best-fit values of $r_0$ and $\gamma$.  The observed
evolution of the correlation lenght $r_0$ is reproduced more
explicitly in the left panel of Fig.~\ref{r0_variance}.  The
interpretation of this diagram in terms of evolution of structure
requires some care.  The flux-limited nature of the survey implies
that the median intrinsic luminosity of each sub-sample at different
redshifts is steadily increasing with redshift.  This means that while
at $z<0.5$ we are essentially measuring the clustering of a population
of low-luminosity galaxies with typical blue absolute magnitude
$M_B\sim -17.5$, in the most distant bin ($z\sim 1.5$), we are
sampling only the very luminous $M_B\sim -21$ galaxies.  We know that
at the current epoch luminous galaxies are more clustered than faint
ones (e.g. Norberg et al. 2001).
In fact, typical $L^*$ galaxies in
the local Universe have a correlation length $r_o\simeq 5 \hmpc$, to
be compared to the value $r_o\simeq 2 \hmpc$ we measure here in our
first redshift bin.  Thus, the ``evolution'' we see in
Fig.~\ref{xi_vvds_1} has little to do with the true evolution of
large-scale structure.

This provides us with a first example of how what we actually measure
is a combination of true physical evolution and observational
selection effects, which are unavoidable in any deep galaxy survey: in
particular, the flux limit and the photometric band in which the
survey is selected, translate, respectively, into a continuously
increasing luminosity limit and a bluer and bluer rest-frame band
sampled as a function of redshift.  Additionally, the pure evolution
of structure is further hidden by the evolution of the relationship
between a given population of galaxies (either selected by luminosity,
color or morphological type) and the underlying mass distribution.
For example, the VVDS has established very clearly that over the
redshift range considered, galaxy stellar populations evolve
strongly, resulting in a brightening of the galaxy 
luminosity function of more than a magnitude 
(Ilbert et al. 2005, Zucca et al. 2006).

So, which galaxies at $z=0$ should correspond to those we are mesuring
at $z=1$?  As far as their clustering is concerned, our ignorance on
how the observed galaxy two-point correlation function at a given
redshift relates to the true $\xi_\rho(r)$ of the mass density can be
parameterized via the {\it bias parameter} as
\be
\xi_g(r,z) = b_L^2(z)\, \xi_\rho(r,z)\,\, .
\label{eq:bias}
\ee 
This expression describes the fact that the galaxy distribution can be
in general a ``biased'' map of the true mass distribution, a concept
which has become familiar in cosmology since the 1980's 
(Kaiser 1984, White et al. 1987).
In this simple form, it assumes that the bias is
scale-independent, which is probably reasonable above a few Mpc
separation.  The linear bias $b_L$ will include all or part of the
effects mentioned above, depending on how the measured correlation
function has been cleaned of the main observational biases. The next
section will present the result of trying and circumvent these effects
and learn more on how they evolve with redshift.

\section{The evolution of galaxy bias}

A more general expression for eq.~(\ref{eq:bias}) can be written in
terms of density contrasts in the galaxy and density fields, 
\be
\delta_g(\delta,z) = b(\delta, z, R)\delta\,\,\, ,
\label{eq:non-linear-bias}
\ee
with $\delta = (\rho - \left<\rho\right>)/\left<\rho\right>$,  implying
that a fluctuation in the galaxy counts can depend on the fluctutation
in the mass in a way that is in general non-linear (i.e. depending on
the density contrast $\delta$ itself), scale-dependent and varying
with redshift.  

\begin{figure}[t!]
\begin{center}
\epsfxsize=6.cm\epsfbox{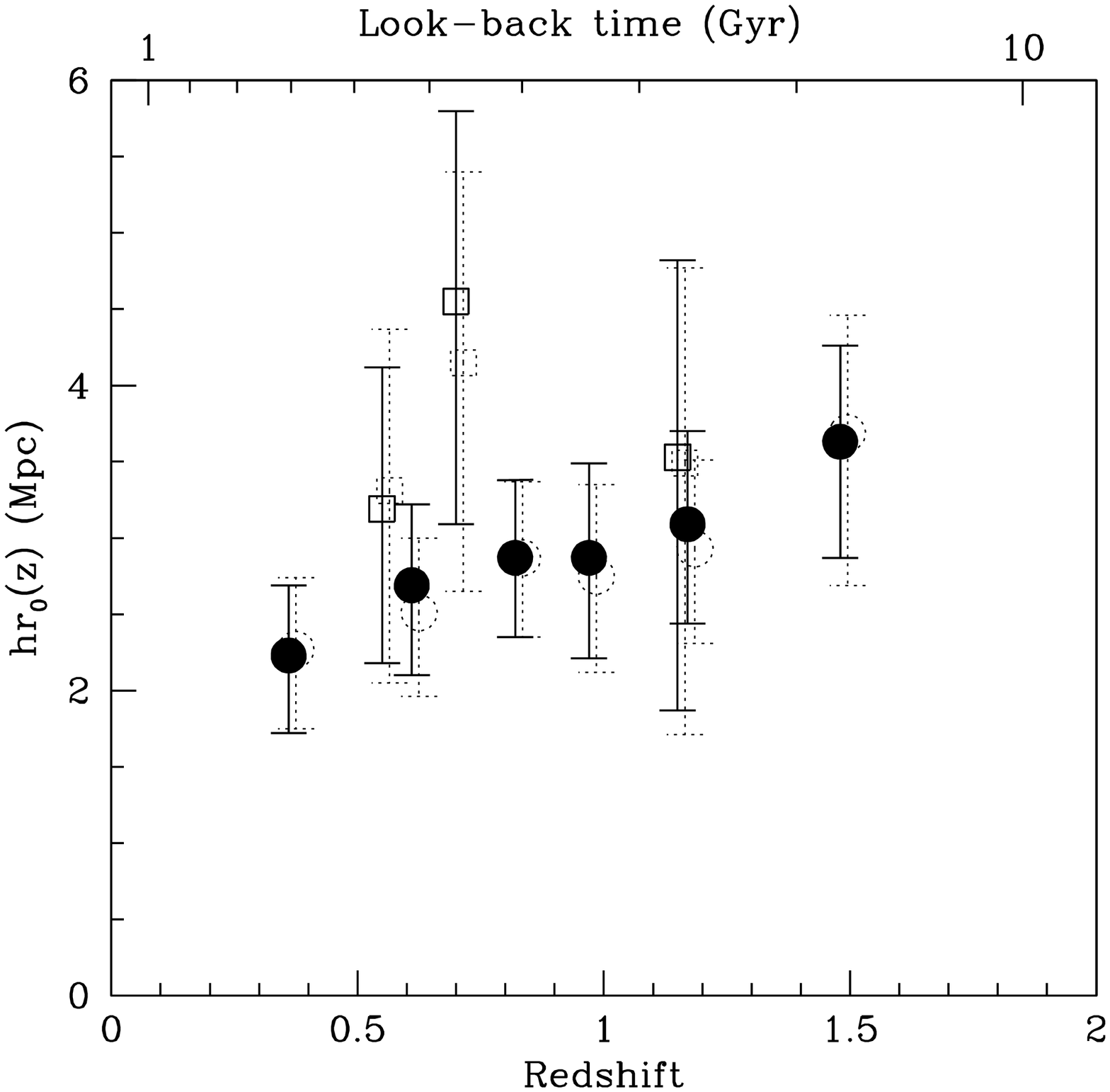}
\epsfxsize=6.cm\epsfbox{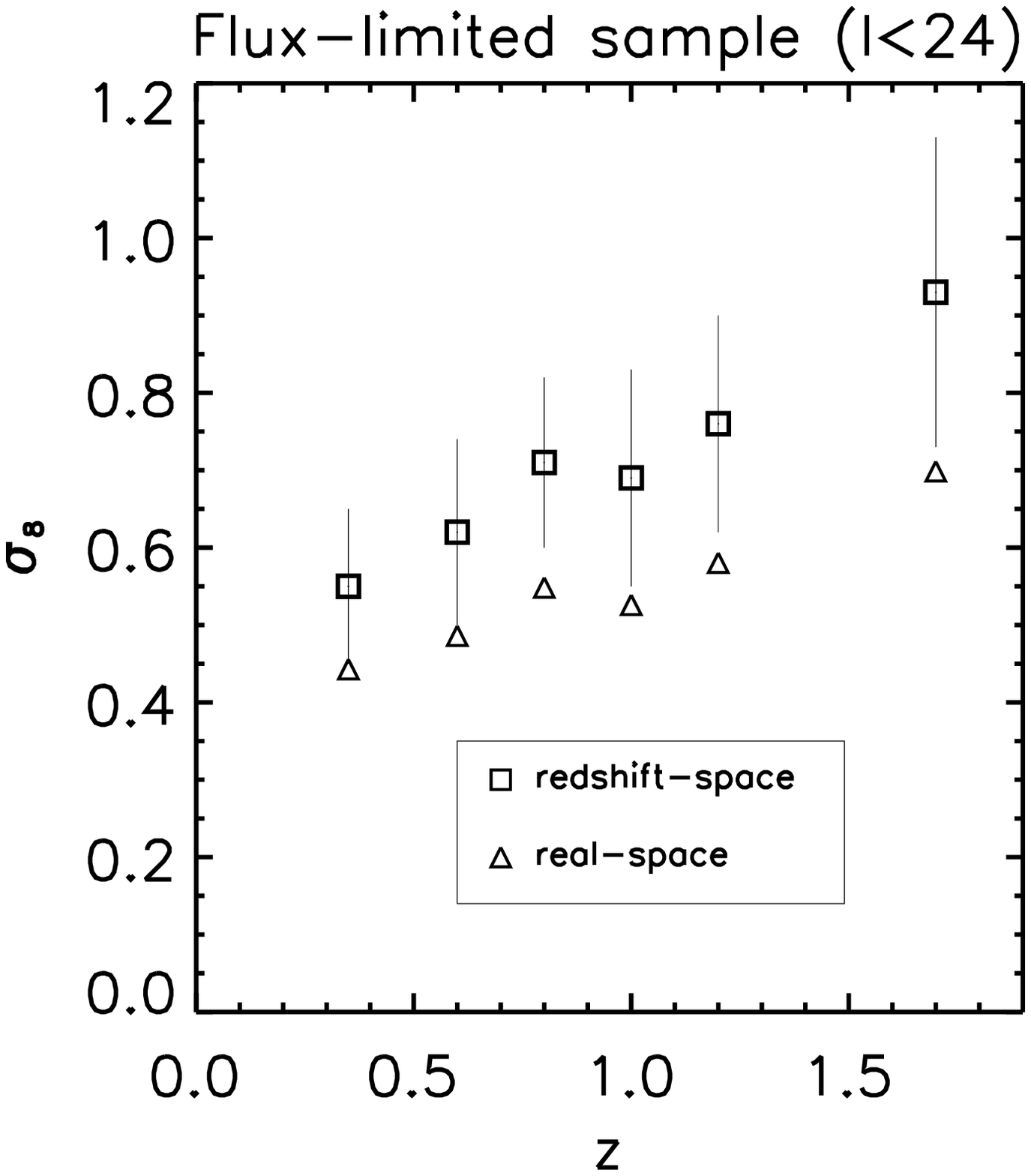} 
\caption{{\it Left:}  Evolution of the correlation length
$r_0$ as a function of redshift using the complete VVDS-Deep flux-limited
data.  Filled circles are from the F02 field of Fig.~\ref{xi_vvds_1},
while squares are from independent measurements in the CDFS, where VVDS-Deep
measured another 1500 redshifts (Le Fevre et al. 2004).  Dashed
symbols show the effect of fixing $\gamma=1.8$ in the fit. 
{\it Right:} The corresponding trend of the standard deviation in galaxy counts
on 8 $\hmpc$ scale, in real (triangles) and redshift (squares) space.
}  
\label{r0_variance}
\end{center}
\end{figure}
The full function $b(\delta, z, R)$ can be estimated from a redshift
survey like the VVDS, by measuring the statistical distribution of
galaxy number counts in spheres of a given size $R$.  Once we assume
the cosmological model, then the corresponding expected probability
distribution function of mass fluctuations $f(\delta)$ can be computed
(e.g. Kayo et al. 2001).  
$b(\delta, z, R)$ can then be obtained, for
a given $R$ and redshift, as the function that maps $f(\delta)$ into
the corresponding function $g(\delta_g)$ obtained from galaxy counts.
The analysis of the VVDS (Marinoni et al. 2005),
shows that
the functional form of the bias is in fact generally more complex than
the simple linear assumption of eq.~\ref{eq:bias}.  
Non-linearity between the galaxy
fluctuation field $\delta_g$ and the matter density field $\delta$ is
detected (for the first time) at a level of $\sim 10\%$.   This result
also shows that the linear approximation of the bias, which is useful and
intuitive in many contexts, can indeed be used within this level of inaccuracy.

The right panel of Fig.~\ref{r0_variance} shows the evolution of the
standard deviation in the VVDS galaxy counts within spheres of $8
\hmpc$, in nearly the same
redshift bins used in the left panel for $\xi(r)$.  The two figures
are essentially equivalent, given the direct relationship between the
variance and the two-point correlation function
\be
\sigma_R^2 = {\frac{1}{2\pi^2}}\int_0^\infty P(k)\, W^2(kR)\, k^2 \,dk \,\,\, .
\ee
\be
P(k) = \int_0^\infty \xi(r) {sin(kr) \over kr}\, r^2 \, dr \,\,\, ,
\ee
where, $W_k$ is the Fourier transform of the filter over which the
observed galaxy distribution is smoothed (in our case, a sphere of
radius $R=8\hmpc$), and $P(k)$ is the power spectrum, the Fourier
transform of $\xi(r)$.

In the figure, the squares refer to the values measured in redshift
space, while the triangles correspond to the real-space $r_0$ and
$\gamma$ obtained via the projected function.  Note how the variance
of the galaxy distribution apparently increases with redshift,
following the trend seen for $r_0$.  Again,
however, this figure is comparing apples with pears, i.e. very faint
galaxies (with small variance) at low redshift, with luminous
objects at $z>1$.

\begin{figure}
\begin{center}
\epsfxsize=6.6cm\epsfbox{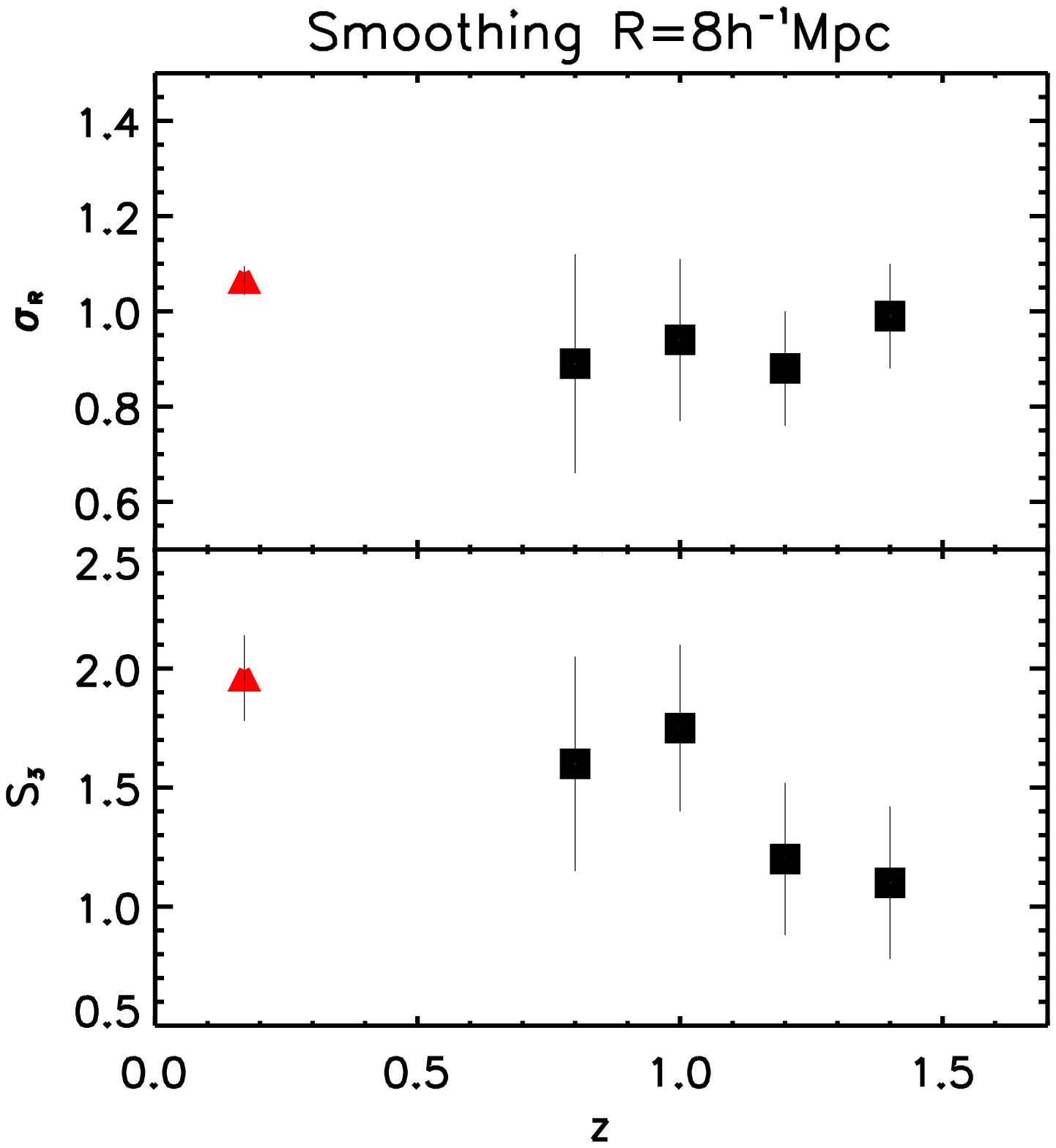}
\epsfxsize=5.7cm\epsfbox{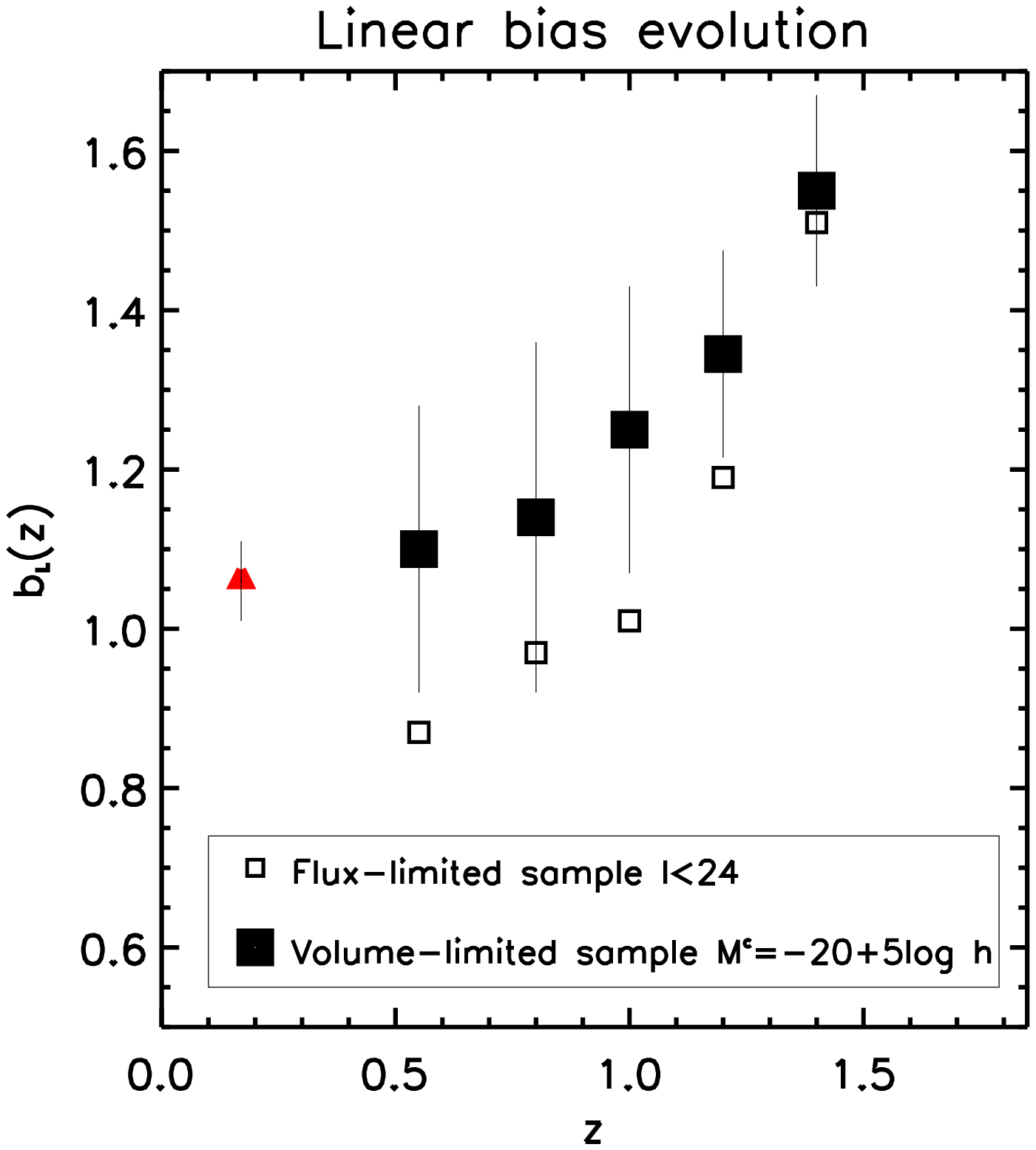}
\caption{ {\it Left:} Evolution of the standard deviation (top) and
  skewness (bottom) of the distribution of galaxy counts in spheres of
  $8 \hmpc$ size, for a sample with $M_B<-20+5\log h$.  Comparison to
  Fig.~\ref{r0_variance}, shows that the observed apparent increase of
  $r_0$ and $\sigma_8$ for the flux-limited sample in fact results
  from the combination of the varying luminosity mix (more and more
  luminous galaxies dominate the redshift bins) with a true increase
  of the bias value as a function of $z$.  {\it Right:} The
  corresponding linear bias evolution (filled squares), compared to
  that from the full magnitude-limited VVDS (open squares, see text
  and ref.$^{33}$ for details).  In all panels, the
  triangle gives the local values from 2dFGRS$^{34}$. }
\label{variance_skewnessM-20} 
\end{center}
\end{figure}

The left panel of Fig.~\ref{variance_skewnessM-20}, shows instead the
result of measuring the second and third moments (standard deviation
and skewness) of the galaxy Probability Distribution Function (PDF) on
scales of 8~h$^{-1}$ Mpc, as a function of $z$, using only galaxies
brighter than $M_B<-20+5\log h$, i.e. in volume-limited
sub-samples.  This involves measuring the number of galaxies in
spheres of 8~h$^{-1}$ Mpc radius and computing the usual statistical
quantities.  The variance remains now constant over the explored
redshift range (while the variance in the matter is expected to
decrease with redshift as a consequence of gravitational growth).  At
the same time, the skewness of the PDF becomes smaller as a function
of redshift, i.e. going back in time. This is a general expectation of
the gravitational instability picture: the growth of fluctuations
modifies the initially Gaussian PDF, skewing it towards positive
values of $\delta_g$.

In practice, the ratio of the values of $\sigma_8$ plotted in
Fig.~\ref{variance_skewnessM-20} (corrected for redshift-space
distortions), to the those expected for the overall dark-matter field
in the adopted cosmology, is the linear bias $b_L$.  As we mentioned
above, from the VVDS Marinoni et al. (2005) 
measured the full bias
function $b(\delta)$ by inverting the relation between the galaxy and
density fields.  An unbiased estimator of the linear bias in this case
is $b_L^2=\left<b^2(\delta)\;\delta^2\right>/\left<\delta^2 \right>$,
which can then be compared to other existing estimates, as we do in
the right panel of Fig.~\ref{variance_skewnessM-20}.  Here, the
evolution of $b_L$, computed at different redshifts both for the full
flux-limited data set (open squares) and for volume-limited samples
with $M_B<-20+5\log h$ (filled squares), is shown.  Note how the
values of the effective bias from the former sample approach
asintotically those from the latter, since at increasing redshift it
becomes more and more dominated by the same luminous galaxies.  This
shows explicitly (note the $z\sim 0.1$ 2dFGRS point) how our
low-redshift measurements of $r_0$ from the full VVDS survey refer to
a population of faint galaxies apparently anti-biased, i.e. more smoothly
distributed than the $\sim L^*$ objects typical of 2dFGRS and SDSS.

Still, also with the volume-limited computation in
Fig.~\ref{variance_skewnessM-20} we are neglecting the important fact,
that also the mean luminosity of galaxies evolves with redshift:
galaxies were brighter on average in the past, with a brightening of
the characteristic luminosity of the luminosity function in this band,
$M_B^*$, of more than 1 magnitude between $z=0$ and
$z=1.4$
(Zucca et al. 2006).  
This means that even with a sample limited to the same
absolute magnitude at all redshifts, we are not selecting strictly the
same population of objects at different distances/epochs.  If we
assume this to be a pure luminosity effect (galaxies become brighter
with redshift, but conserve their total number in comoving
coordinates), this implies that the mass-luminosity ratio in the
observed band decreases as a function of redshift.  As a consequence,
by applying a cut at constant luminosity $M_B<-20$, we are including
at increasing redshifts galaxies of smaller and smaller mass.  Thus,
had one used an evolving limit, e.g. by selecting galaxies brighter
than a fixed $L_{lim}(z)/L^*(z)$ (i.e. a constant
$M_{lim}(z)-M^*(z)$), the resulting ``true'' (i.e. mass-related) bias
evolution would be even steeper.  We have used this approach to study
the dependence of clustering on absolute luminosity at $z\sim 1$ in
the VVDS and compare it to local results, as discussed in the next
section.

\section{Clustering of galaxies with different luminosity at $z=1$}
\label{xi_lum}

At the current epoch, luminous galaxies are observed to be more
clustered than faint ones (e.g.Norberg et al. 2001 and references
therein), with this difference becoming more significant above the
characteristic luminosity $L^*$.  This effect is in general agreement
with the predictions from hierarchical models (e.g. Benson et
al. 2001), in which bright galaxies are expected to form in more
massive dark matter halos, which are typically more strongly clustered
than the overall distribution of dark matter 
(Kaiser 1984, Mo \& White 1996).
In this scenario, these differences should be enhanced
at high redshifts, where galaxy formation is expected to be more and
more confined to the highest peaks of the density field.

We have addressed this issue with the VVDS, by measuring the
correlation functions $\xi(r_p,\pi)$ and $w_p(r_p)$ for a series of
sub-samples with increasing median luminosity $M_B$ comprised between
$-19.7$ and $-21.3$  and covering the redshift range [0.5-1.2] (median
redshift $z\simeq 0.9$) (Pollo et al. 2006).
%
\begin{figure}
\begin{center}
\epsfxsize=6.2cm\epsfbox{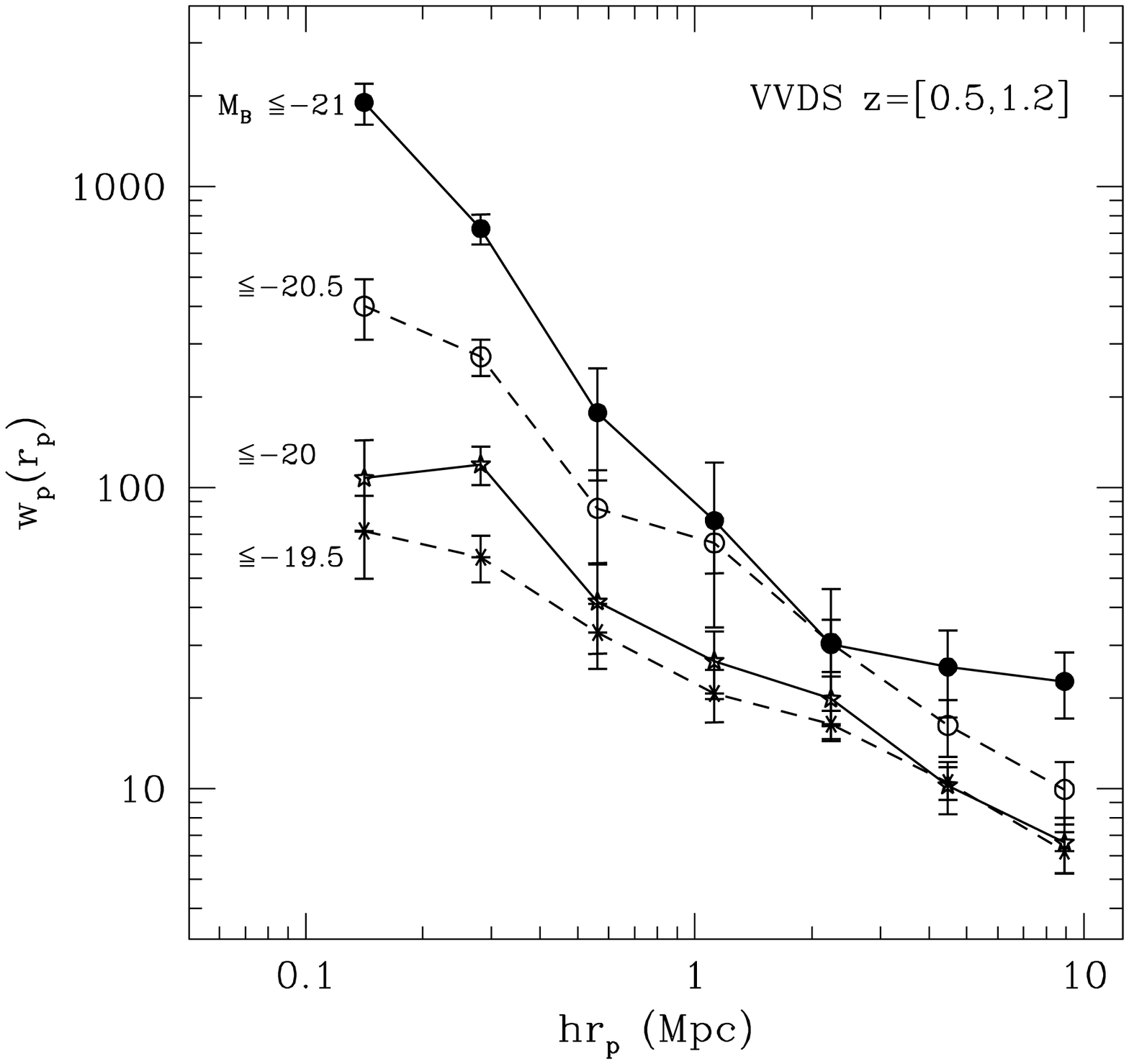}
\epsfxsize=6.2cm\epsfbox{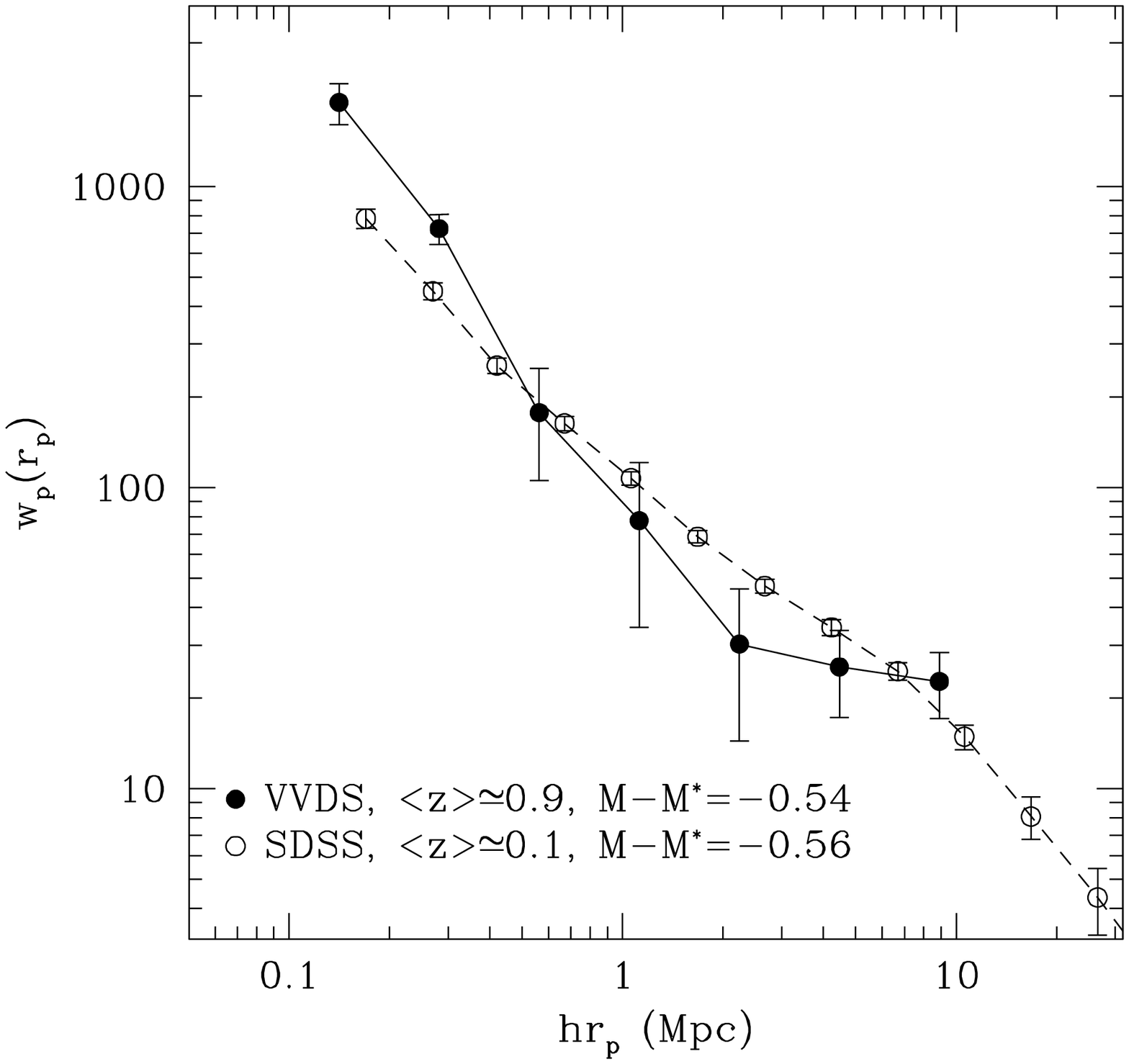} 
\caption{{\it Left:} Projected correlation function $w_p(r_p)$ from
four high-redshift volume-limited sub-samples of the VVDS with
increasing luminosity.  {\it Right:} Comparison of the SDSS
and VVDS measurements for samples with comparable luminosity (relative
to $L^*$) respectively at $z\simeq 0.1$ and $z\simeq 0.9$.
}
\label{wp_lum} 
\end{center}
\end{figure}
The left panel of Fig.~\ref{wp_lum} shows the projected function for
the four brightest volume-limited high-redshift sub-samples of the
VVDS.  The right panel instead makes a direct comparison of the shape
of $w_p(r_p)$ for galaxies with the same relative luminosity ($\sim
0.5$ magnitudes brighter than $M_B^*$) in the SDSS at $z\sim 0.1$ and
in the VVDS at $z\sim 0.9$.  This trick of expressing the typical
luminosity of a sample relatively to the characteristic luminosity
$L^*$ at the same epoch is an attempt to factor out the overall
luminosity evolution,
thus hopefully comparing galaxies of similar mass at different
redshifts.  Two main effects are evident from these figures.  On one
side, the clustering amplitude does depend on luminosity also at
redshifts of the order of unity.  Additionally, however, we observe a 
steepening of the small-scale part of $w_p(r_p)$ for highly-luminous galaxies,
significantly stronger than in local SDSS samples (Zehavi et al. 2005).


%
\begin{figure}
\begin{center}
\epsfxsize=5cm\epsfbox{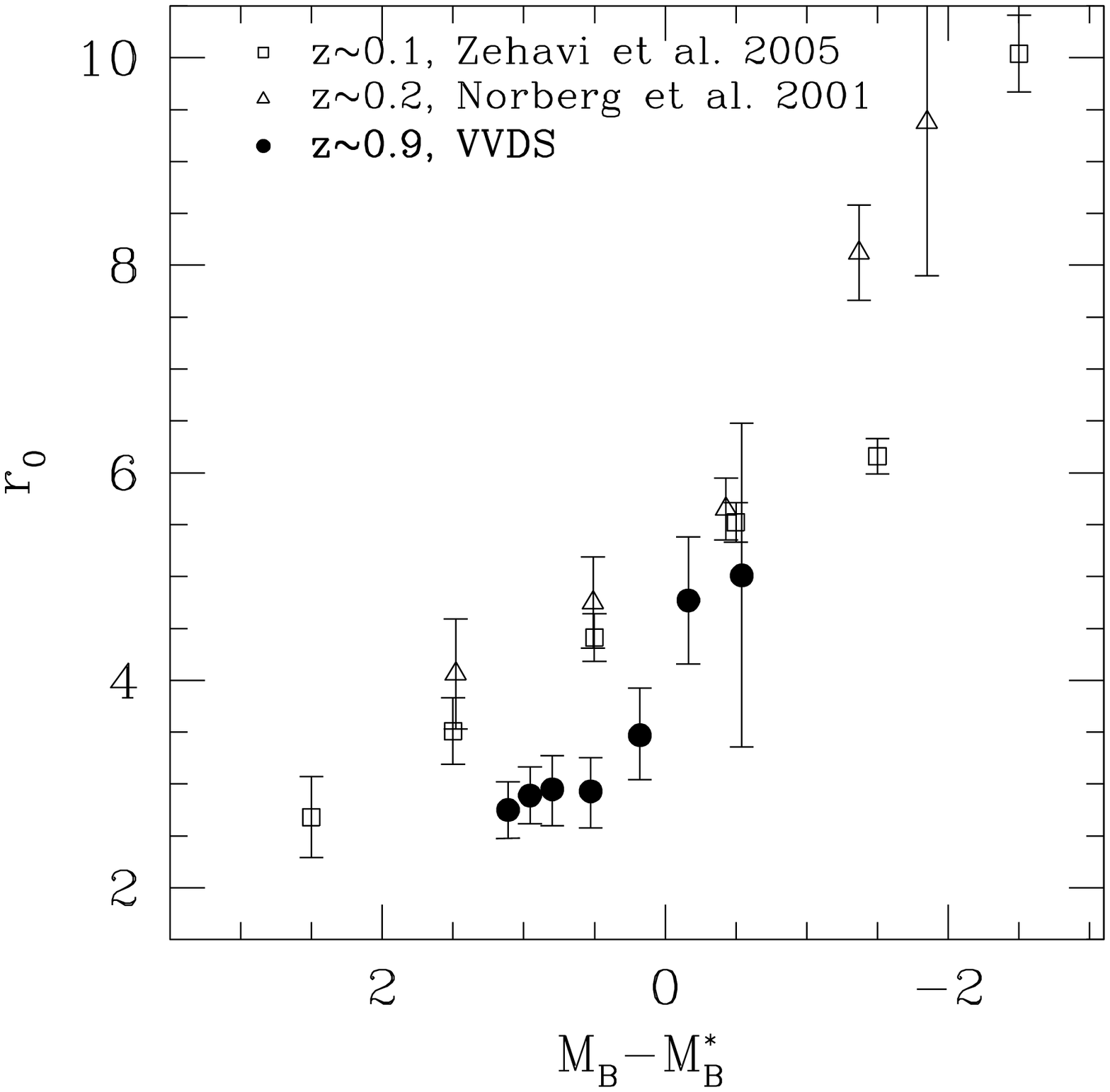} 
\epsfxsize=5cm\epsfbox{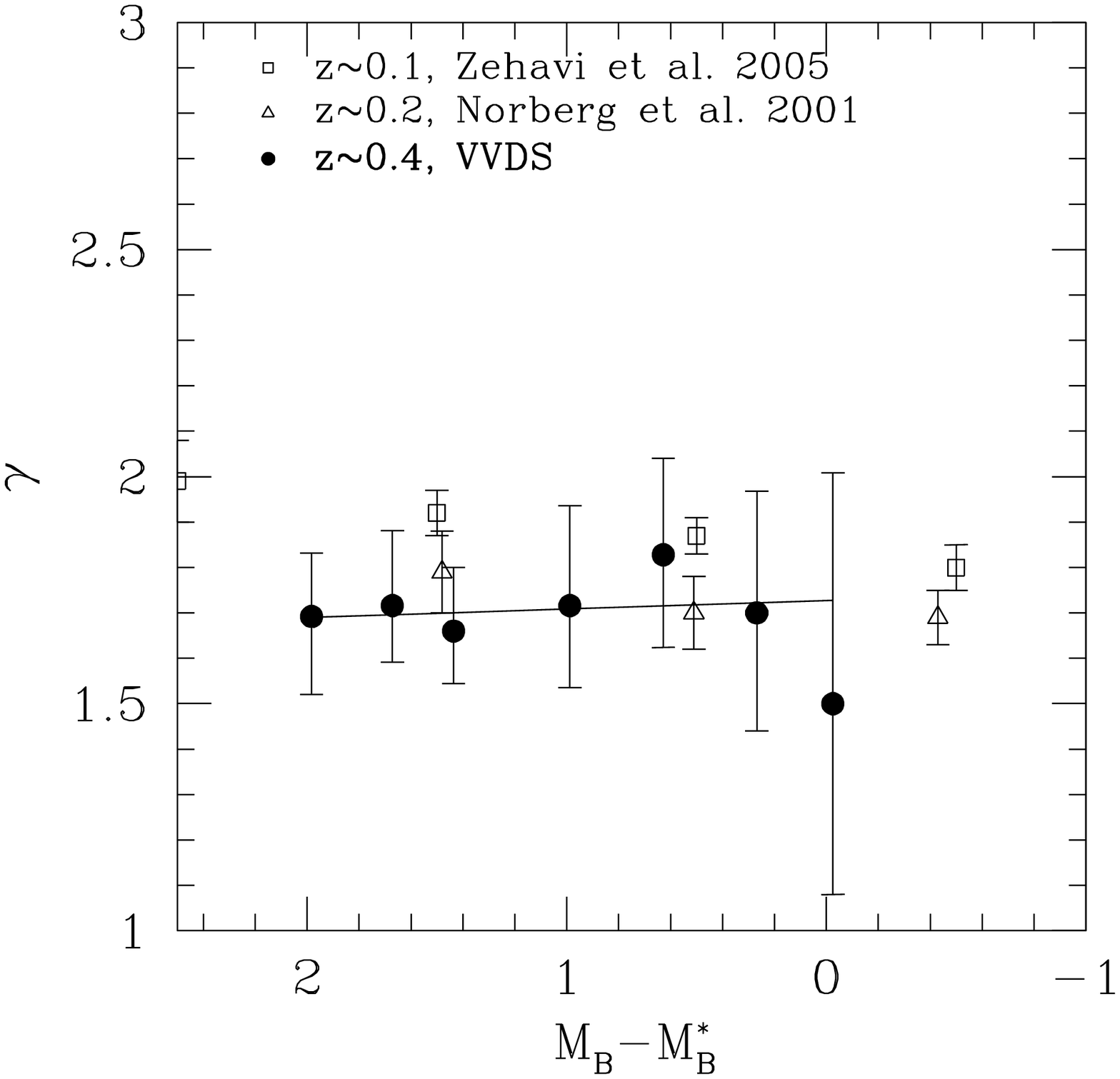} 
\epsfxsize=5cm\epsfbox{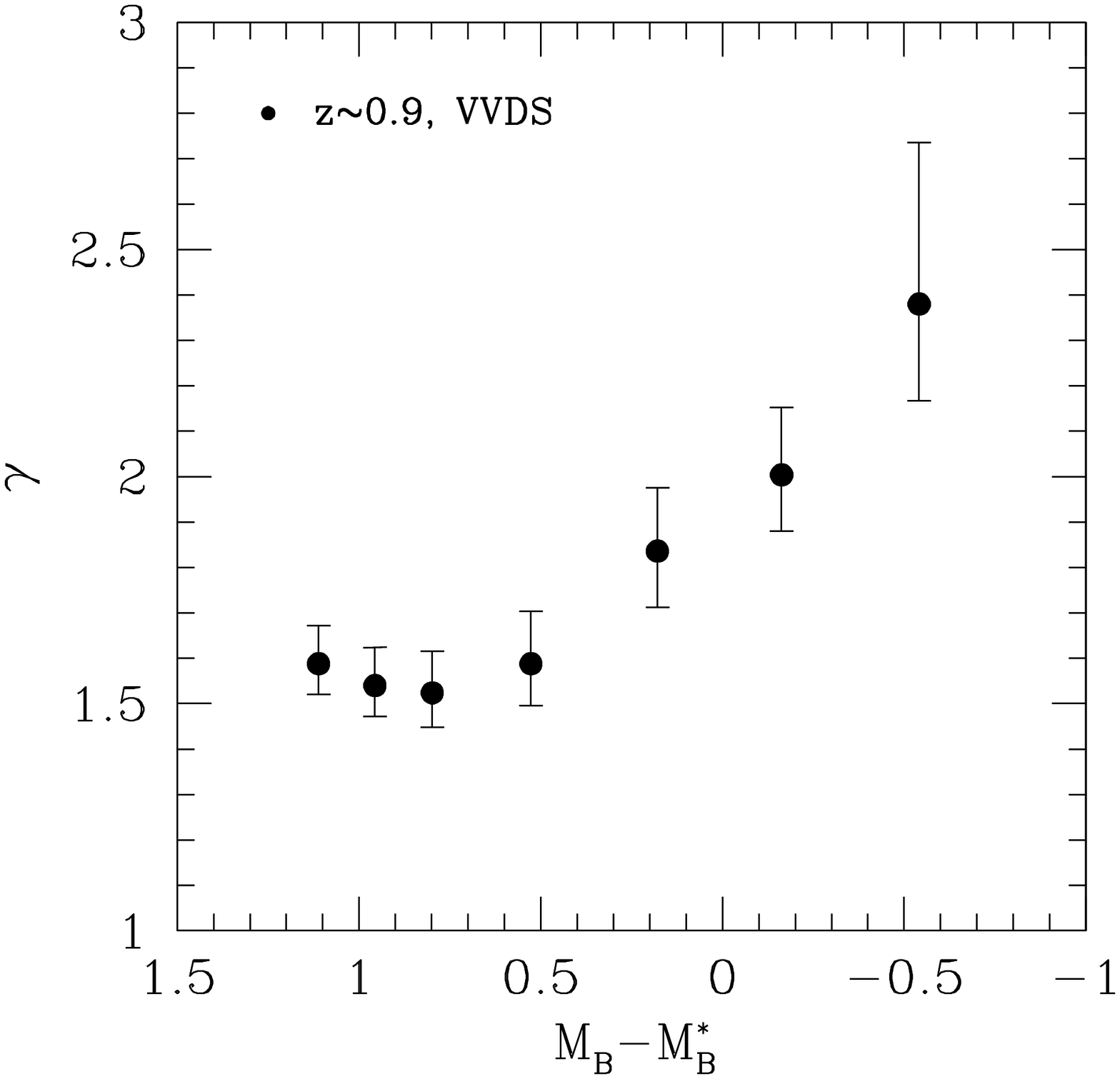} 
\caption{{\it Left:} The dependence of the clustering length $r_0$ on
  galaxy luminosity from 2dFGRS and SDSS compared to the VVDS
  measurements at $z \sim 0.9$.  {\it Center:} Values of the slope of
  $\xi(r)$, $\gamma$, as a function of luminosity at low redshift .
  {\it Right:} The same, but at $z\sim 0.9$.  Note the
  dramatic steepening of $\xi(r)$ at increasing luminosities in this
  redshift range.  }
\label{comp} 
\end{center}
\end{figure}
%
%
This effect can be seen even more explicitly in Fig.~\ref{comp}, where
we compare our measurements of $r_0$ and $\gamma$ from the VVDS
samples to the local values from 2dFGRS (Norberg et al. 2001)
and SDSS (Zehavi et al. 2005):
galaxies fainter than $M_B^*$ at high redshift are
significantly less clustered than their counterparts in the
present-day Universe
(left panel).  At the same time, the clustering strength of galaxies
brighter than $M_B^*$ is comparable to that observed locally with a
correlation length up to $r_0=4.77\pm0.61$ h$^{-1}$ Mpc. We therefore
observe that at redshift $z\simeq0.9$, as luminosity increases above
$L^*$, the clustering length suddenly rises to values comparable to
those observed locally for galaxies with similar $M_B-M_B^*$.  It is
the slope $\gamma$, however, that shows the most dramatic difference
when comparing the $z\sim 0$ and $z\sim 1$ results, and from
Fig.~\ref{wp_lum} we see that this change is mostly confined to scales
$r_p<2 \hmpc$.  A parallel, analogous work on the DEEP2 data (Coil et
al. 2006) finds a similar result, with slightly less pronounced
steepening.  Even more interestingly, Ouchi et al. (2005) observe the
same phenomenon studying the clustering of Lyman-break selected
galaxies at $z\sim 4$.  In this analysis, they also find that it is
the small-scale part of the correlation function which is the most
sensitive to the mean luminosity of the sample, steepening for more
luminous galaxies as in the case of the VVDS at $z\sim 1$. 

Although with a much less evident steepening (note that the slope in
the central panel of Fig~\ref{comp} is measured between 0.1 and 10
h$^{-1}$ Mpc, thus weakening any small-scale effect),
the existence of a feature in $\xi(r)$ on scales $\sim 1-3$ $\hmpc$ is
in fact a well-established observation also for luminous galaxies in
the local Universe.  First evidences go back to analyses of the
pioneering redshift surveys of 1980's, as the CfA and Perseus-Pisces
surveys (Dekel \& Aarseth 1984; Guzzo et al. 1991),
where quite naturally this feature was interpreted
as marking the transition scale between the linear regime of clustering
on large scales and fully nonlinear structures on small scales. This
picture was found to be consistent with both numerical 
(Branchini et al. 1994)
and analytical (Peacock 1997)
results on the non-linear evolution of phenomenological scale-free or
CDM-like power spectra.  The same effect was also particularly evident
in the power spectrum obtained from deprojection of the APM angular
correlation function (Baugh \& Efstathiou 1993, see also Guzzo 1997
and Gazta\~naga \& Juszkiewicz 2001),
and in the reconstructed shape of the general galaxy power spectrum
(Peacock \& Dodds 1994).  

The reality of this feature in the galaxy correlation function has
been confirmed to high accuracy in more recent times by the SDSS data
(Zehavi et al. 2004) 
and interpreted in the context of {\sl halo occupation distribution
  (HOD)} models.  In these models, a statistically motivated recipe to
describe galaxy formation determines the halo occupation distribution,
specifying the probability $P(N|M)$ that a dark matter halo of virial
mass $M$ contains $N$ galaxies of a given type.  This term (known as
the {\it one-halo component} of the correlation function) governs the
behaviour of galaxy correlations on small ($<2$ h$^{-1}$ Mpc) scales
(i.e. where $w_p(r_p)$ steepens), while at larger separations galaxy
correlations are described by the quasi-linear clustering of the halos
(the {\it two-halo component}), with essentially no dependence on the
more complex physics of the sub-dominant baryonic component (see
Cooray \& Sheth 2002
for a comprehensive review).  The modern HOD scenario, therefore, confirms
the interpretation of the small-scale steepening of galaxy
correlations as the transition to fully non-linear clustering, as
speculated in Guzzo et al. (1991), but specifying it as the scale
below which correlations are dominated by pairs of galaxies belonging
to the same dark-matter halo.

HOD models are found to provide a
better description of the non-linear clustering of galaxies with
respect to analytical scaling formulae (as e.g. the remarkable one by
Hamilton et al. 1991), used in earlier times to predict the observed
non-linear shape of the galaxy power spectrum or correlation function
(see e.g. Smith et al. 2003).   Still, 
the number of free parameters that need to be
constrained by comparison to observed properties (as e.g. the
dependence of clustering on galaxy luminosity or color), make the
overall technique not fully satisfactory.  Recently, there have been
interesting attempts to go beyond this substantially statistical
description of what happens to galaxies on sub-halo scales (Conroy et
al. 2006; Wang et al. 2006). These works try to improve the definition
of galaxies within dark-matter halos identified in N-body simulations,
in particular by accounting for objects whose dark-matter halos are destroyed
by tidal effects, but that still survive in their baryonic component.
Conroy et al. (2006) specifically study the dependence of the shape of
the two-point correlation function on luminosity at different
redshifts.  Remarkably, they correctly reproduce the small-scale
upturn of the correlation function in luminosity-selected samples and
also show (as we have found from our comparison of the VVDS and SDSS
data), that this deviation is stronger at higher redshifts and for
more luminous objects.  

Along these lines, work is in progress to perform accurate comparisons of VVDS
to mock galaxy surveys from the Millennium Simulation.

\section{The clustering of different galaxy types up to $z\sim
1.5$}\label{sec:xi_types}

It is quite plausible that the relationship between galaxies and their
parent dark-matter halos depend on the galaxy type or, better, that
the physical morphological properties of galaxies depend in some way
on the properties of the hosting halo, as e.g. its mass.  In fact, it
is a well-established observational fact in the local Universe that
red galaxies are more clustered than blue galaxies, which at least at
$z=0$ is equivalent to say that early-type (elliptical and S0's)
galaxies are more strongly packed among themselves than late-type
(spiral and irregular) ones (e.g. Guzzo et al. 1997 
and references therein).  
In the now familiar bias jargon
introduced in the previous sections, we may rephrase this by saying
that red galaxies are a {\it more biased} population than blue
galaxies, and thus a less faithful tracer of the true mass
distribution.  Indeed, elliptical and S0 galaxies are the preferred
population of rich galaxy clusters, giving rise to the well known
morphology-density relation (e.g. Dressler 1980).

It is thus natural to ask: is this difference already established at
redshifts $z\sim 1$?  And is the way this difference evolves with
redshifts telling us anything about the way galaxies in general trace
the underlying structure evolution?  One possibility, for example, is
that old massive elliptical galaxies are simply passively evolving
within this redshift range 
(e.g. Cimatti et al. 2006), 
and thus their clustering should follow almost passively the growth of
structure.  The alternative scenario is that of a significant merging
activity since $z\sim 1$ to today (e.g. Bell et al. 2004),
that should manifest itself in a different evolution of
the clustering of this population.
\begin{figure}
\begin{center}
\epsfxsize=6cm\epsfbox{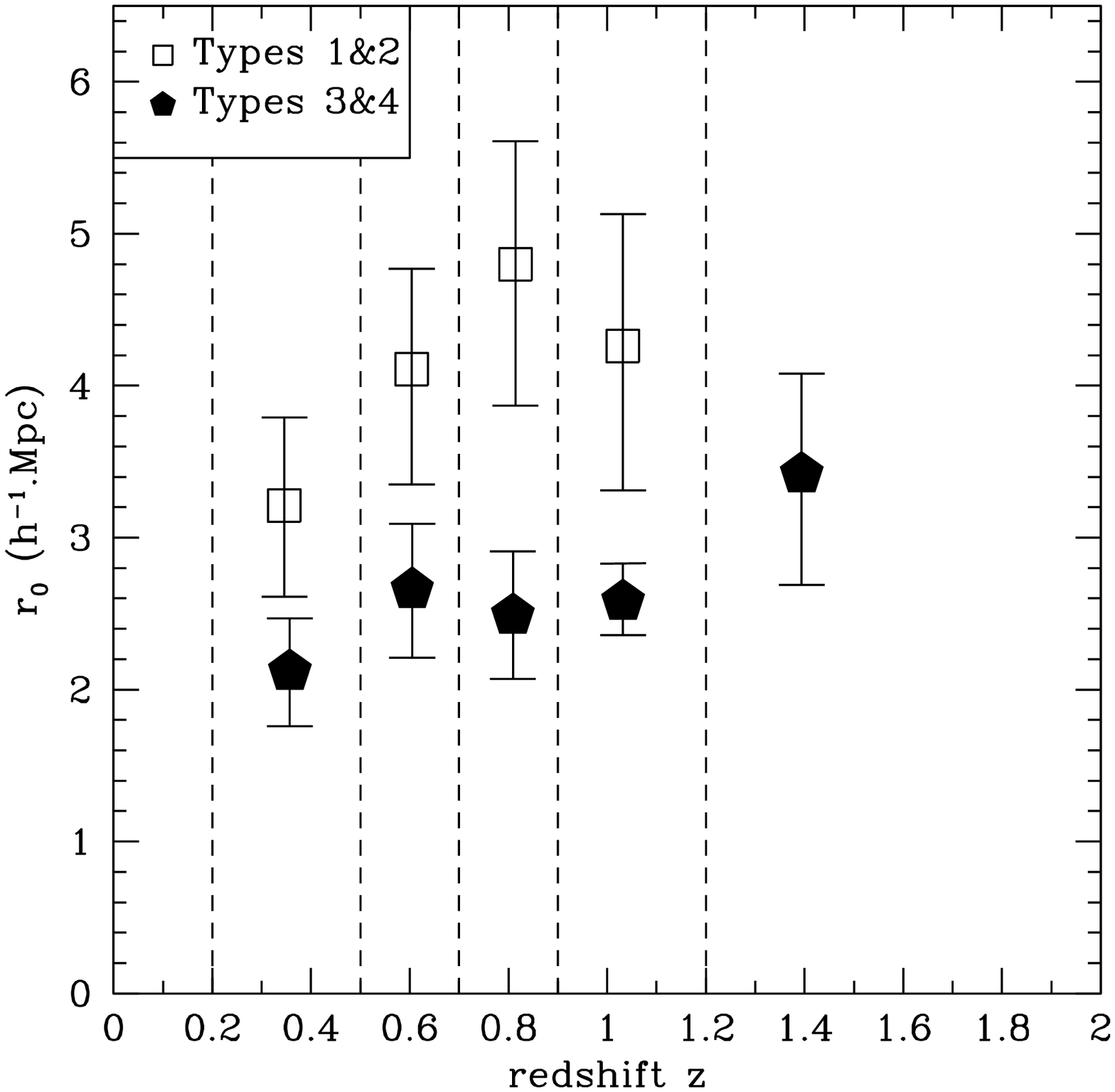}
\epsfxsize=6.3cm\epsfbox{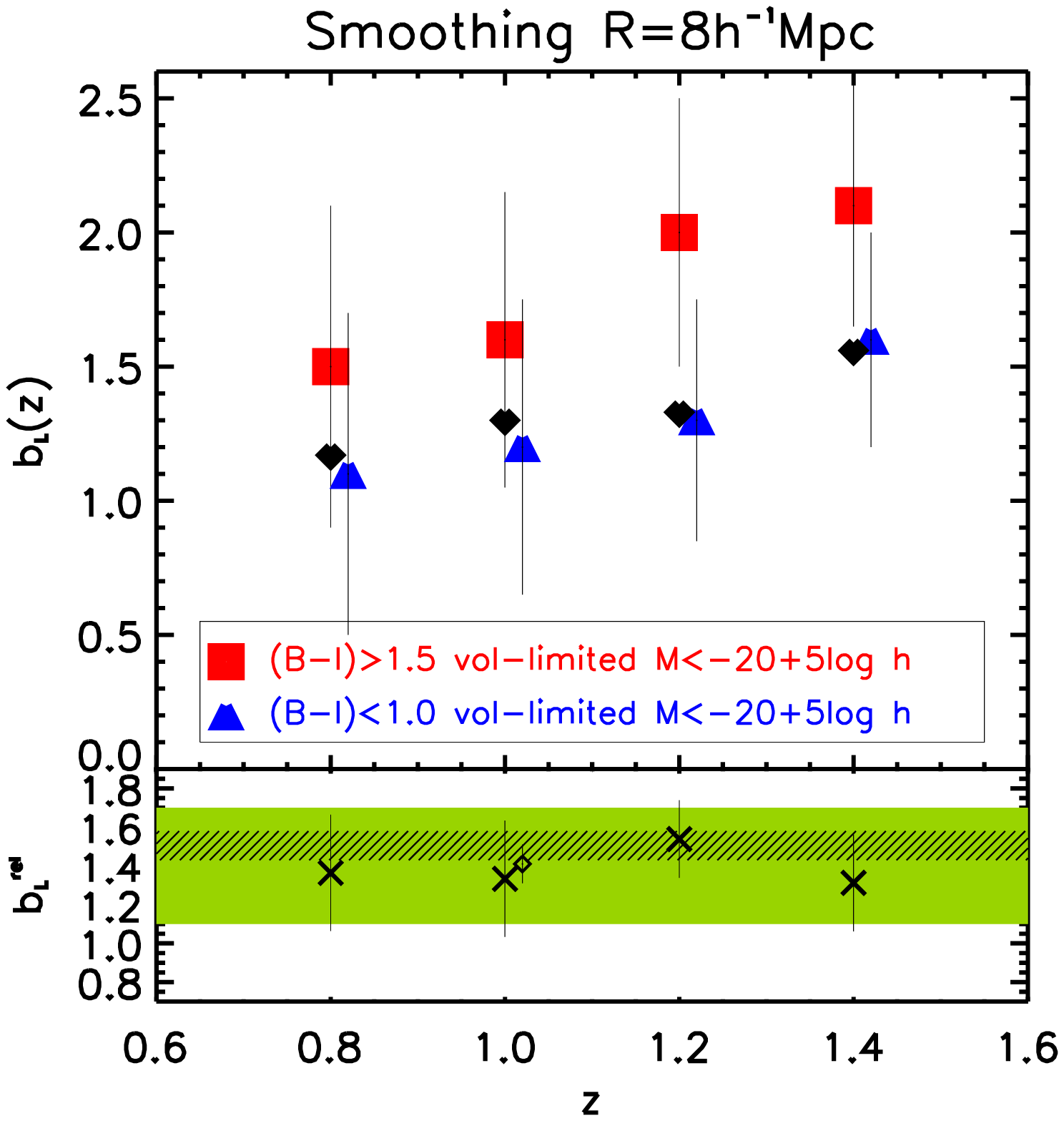}
\caption{{\it Left:} Evolution of $r_0$
  for red (open symbols, spectral types $1+2$) and blue (filled
  symbols, types $3+4$) galaxies from the whole VVDS (Meneux et
  al. 2006).  
  {\it Right:} The linear bias of red (squares) and blue (triangles)
  galaxies with respect to dark matter, computed as discussed in the
  text, for samples with $M_B<-20+5\log h$. Small filled lozenges show
  the value for all types, corresponding to the right panel of
  Fig.~\ref{variance_skewnessM-20}.  The bottom panel gives
  the relative bias of the two classes, that remains
  constant over this range.  }
\label{xi_types} 
\end{center}
\end{figure}

We have addressed this issue with the VVDS data, and the results are
presented in the recent paper by Meneux et al. (2006).
Galaxies have been split into four spectral classes, according to a
best-fitting template procedure to their multi-band photometry, as
discussed in detail in Zucca et al. (2006).
The left panel of Fig.~\ref{xi_types} presents the evolution of the
correlation lenght $r_o$ between $z\sim 0.3$ and $z\sim 1.5$, for
early- and late-type galaxies separately.  The main result from this
plot is that at least out to $z\sim 1.2$ red galaxies remain more
clustered than blue galaxies.

Being obtained from the full flux-limited survey, this plot includes
the same luminosity effect discussed in section~\ref{xi_1}, i.e. it
compares low-luminosity objects at $z\sim 0.3$ with the most luminous
galaxies at $z > 1$.  In fact, we know from local samples
(e.g. Guzzo et al. 1997) that bright $M_B \sim M_B^* \simeq -19.5$
galaxies at $z\sim 0$ show $r_o\simeq 8 \hmpc$ and $r_o\simeq 5 \hmpc$ for
early- and late-types respectively.  The plot shown in the right panel
of Fig.~\ref{xi_types} Marinoni et al. 2005),
again tries to
overcome this problem, at the expense of reducing the sample
significantly by selecting -- at any redshift -- only galaxies
brighter than a fixed intrinsic luminosity.  Here the relative
strength of clustering of early and late spectral types is expressed
in terms of their linear bias, computed as in
Fig.~\ref{variance_skewnessM-20} (see 
Marinoni et al. 2005 
for details).  The net result is that out to $z\sim 1.5$ the relative
bias between red and blue galaxies remains practically
constant, as shown explicitly in the bottom part of the figure.  This
seems apparently to contraddict the recent measurement from the same
VVDS data of a clear evolution of the color-density relation.  As
discussed in detail in the recent paper by Cucciati et al. (2006),
the VVDS data show that the color-density relation weakens
significantly above $z\sim 1$, suggesting that star-forming galaxies
start to populate more and more the high-density regions.
  
The difference with Fig.~\ref{xi_types} might perhaps be just the
consequence of different sensitivity of the first and second moments
(density and variance) to changes in the global environment: finding
blue or red galaxies in high-density regions at $z\sim 1.4$ becomes
equally probable, but still the number of small-scale pairs is larger
for red objects.  On the other hand, the higher-redshift point for red
galaxies in the right panel of Fig.~\ref{xi_types} is also affected by
small statistics (and in fact there is no equivalent point at $z=1.4$
in the left panel, where a differential quantity -- the correlation
function -- is used).  Additional work which is in progress using
Spitzer photometry from the Swire survey (Lonsdale et al. 2003), is in fact
indicating a similar clustering level at $z\sim 1.5$ for luminous,
blue-selected star-forming galaxies and NIR-selected, massive objects
(de la Torre et al. 2007), thus in agreement with the trend indicated
by the color-density relation.
This is one of several investigations on the relationship between
large-scale structure and galaxy properties, that are ongoing with the
VVDS data.  Further boost to these researches will come from the
completion of the redshift measurements for the $\sim 33,000$ spectra
collected over the {\it VVDS-Wide} fields (Garilli et al. 2007)
and the combined use of the {\it VVDS-Deep} spectroscopic sample with
the high-quality photometric redshifts obtained in the surrounding
area using the additional CFHT-LS photometry (Ilbert et al. 2006).
In addition to extending and improving the results obtained from {\it
  VVDS-Deep}, cosmological tests requiring sampling large volumes and
scales $\gg 10$ h$^{-1}$ Mpc will become feasible for the first time
at $z\sim 1$ thanks to the large area coverage provided by the {\it
  VVDS-Wide} data.  These include, for example, the measurement of
$\beta\simeq \Omega_M^{0.6}/b$ from the the distortions of
$\xi(r_p,\pi)$.  Realistic simulations show that with the planned full {\it
  Wide} survey ($\sim 100,000$ redshifts over $\sim 10$ square
degrees), $\beta$ can be measured at $z\sim 1$ to an accuracy of 10\%
(Pierleoni et al., 2007),
thus providing an estimate of the mean matter density when the
Universe was about half its current age. 




%
%



\section*{Acknowledgments}
LG thanks the organizers of the Moriond 2006 workshop for kindly
inviting him to present this review and J. Blaizot, G. De Lucia,
G. Kauffmann and S. White for stimulating discussions on galaxy
clustering and HOD models.  ESO and the Max-Planck Insitut f\"ur
Astrophysik are also gratefully acknowledged for their hospitality and
for providing a stimulating atmosphere during the completion of this
paper.



\section*{References}

{\small
\begin{thebibliography}{99}




\bibitem{SDSS06}Adelman-McCarthy, J.K., et al., ApJS, {\bf 162}, 38 (2006)
\bibitem{Baugh_Efstathiou93} Baugh, C.M., Efsthathiou, G.,
MNRAS {\bf 265}, 145 (1993)
\bibitem{Bell04}Bell, E., et al., ApJ {\bf 608}, 752 (2004)
\bibitem{benson} Benson, A. J., Cole, S., Frenk,
  C. S., Baugh, C.M., Lacey, C.G., MNRAS {\bf 327}, 1041 (2001)
\bibitem{Blaizot_GALICS} Blaizot, J., et al., MNRAS {\bf 360}, 159  (2005)
\bibitem{BG2001} Borgani S. \& Guzzo L., {Nature} {\bf 409}, 39 (2001)
\bibitem{BorganiRDCS2001}Borgani, S., et al., ApJ {\bf 561}, 13 
 (2001)
\bibitem{Branchini94} Branchini, E., Guzzo, L., Valdarnini, R.,
ApJ {\bf 424}, L5 (1994)
\bibitem{Ellis_Broadhurst}Broadhurst, T.J., Ellis, R.S., Shanks, T.,
 MNRAS {\bf 235}, 827  (1988)
\bibitem{Cimatti06} Cimatti, A., Daddi, E., Renzini, A., AA 
{\bf 453}, L29 (2006)
\bibitem{Coil2006}Coil, A.L., et al. ApJ {\bf 644}, 671 (2006)
\bibitem{Cole_2dF}Cole, S., et al., MNRAS {\bf 362}, 505 (2005)
\bibitem{2dF_release} Colless, M., et al., MNRAS {\bf 328}, 1039 (2001)
\bibitem{Conroy06}Conroy, C., Wechsler, R., Kravtsov, A.V., ApJ
647, {\bf 201} (2006)
\bibitem{Cooray_Sheth02} Cooray, A. \& Sheth,
R., Phys. Rept., {\bf 372}, 1  (2002)
\bibitem{Croton04} Croton, D.J., et al., MNRAS {\bf 352}, 1232 (2004)
\bibitem{Cucciati06} Cucciati, O., Iovino, A., Marinoni, C., et
  al. (VVDS Consortium), AA {\bf 458}, 39 (2006)
\bibitem{DP83} Davis, M., Peebles, P.J.E., ApJ {\bf 267}, 456 (1983)
\bibitem{Dekel_Aarseth84} Dekel, A., Aarseth, S., ApJ {\bf 283}, 1
  (1984)
\bibitem{delatorre07} de la Torre, S., et al. (VVDS Consortium), AA,
  submitted 
\bibitem{Dressler80} Dressler, A., ApJ {\bf 236}, 351 (1980)
\bibitem{Eisenstein_BAO}Eisenstein, D., et al., ApJ {\bf 633}, 560 (2005)
\bibitem{Garilli07} Garilli, B., et al. (VVDS Consortium), in
  preparation (2007)
\bibitem{GaztaJuszk} Gazta\~naga, E., Juszkiewicz, R., ApJ {\bf 558},
L1 (2001)
\bibitem{Guzzo91} Guzzo, L., et al., ApJ {\bf 382}, L5 (1991)
\bibitem{GuzzoNewA}Guzzo, L., NewA, {\bf 2}, 517 (1997)
\bibitem{Guzzo97} Guzzo, L., et al., ApJ {\bf 489}, 37 (1997)
\bibitem{Hamilton91}Hamilton, A.J.S., Kumar, P., Lu, E., 
Matthews, A., ApJ {\bf 374}, L1 (1991)
\bibitem{Ilbert_VVDS_LF} Ilbert, O., Tresse, L., Zucca, E., et
  al. (VVDS Consortium), AA {\bf 439}, 863 (2005)
\bibitem{Ilbert06} Ilbert, O., Arnouts, S., McCracken, H.J., AA
  {\bf 457}, 841 (2006)
\bibitem{Jenkins01}Jenkins, A., et al., MNRAS {\bf 321}, 372
\bibitem{Kaiser87_beta} Kaiser, N., MNRAS {\bf 227}, 1 (1987)
\bibitem{Kaiser84} Kaiser, N., ApJ {\bf 284}, L9 (1984)
\bibitem{Kauff_SAM} Kauffmann, G., et al., MNNRAS {\bf 307}, 529 (1999)
\bibitem{Kayo01} Kayo, I., Taruya, A., Suto, Y., ApJ {\bf 561}, 22 (2001)
\bibitem{Lacey_SAM} Lacey, C., Cole, S., MNRAS {\bf 262}, 627 (1993)
\bibitem{LeFevre_CFRS}Le F\`evre, O., et al., ApJ {\bf 461}, 534
  (1996)
\bibitem{Swire} Lonsdale, C.J., Smith, H.E., Rowan-Robinson, M.J., et al.,
  PASP {\bf 115}, 897 (2003)
\bibitem{OLF_VVDS_CDFS} Le F\`evre, O., Vettolani, G., Paltani, S.,
  et al. (VVDS Consortium), AA {\bf 328}, 1043 (2004)
\bibitem{OLF_VVDS_survey1} Le F\`evre, O., Vettolani, P., Garilli, B.,
  et al. (VVDS Consortium), AA {\bf 439}, 845 (2005)
\bibitem{OLF_VVDS_xi1} Le F\`evre, O., Guzzo, L., Meneux., B. et
al. (VVDS Consortium), AA {\bf 439}, 877 (2005)
\bibitem{Lilly2007}Lilly, S., et al. (the zCOSMOS Team), ApJS COSMOS
  Special Issue,  (2007) in press 
\bibitem{Marinoni_VVDS_bias}  Marinoni, C., Le F\`evre, O., Meneux,
B. et al. (VVDS Consortium), AA {\bf 442}, 801 (2005)
\bibitem{Meneux06} Meneux, B., Le F\`evre, O., Guzzo, L., et
al. (VVDS Consortium), AA {\bf 452}, 387 (2006)
\bibitem{Mo_White96}Mo, H.J., \& White, S.D.M., MNRAS {\bf 282}, 347 (1996)
\bibitem{norberg}Norberg, P., Baugh, C.M.,
Hawkins, E. et al, MNRAS {\bf 328}, 64 (2001)
\bibitem{Ouchi2005} Ouchi, M., ApJ {\bf 635}, L117 (2005)
\bibitem{Peacock97} Peacock, J.A., MNRAS {\bf 284}, 885 (1997)
\bibitem{Peacock_Dodds94} Peacock, J.A., Dodds, S.J., MNRAS {\bf 267},
  1020 (1994)
\bibitem{Pierleoni07} Pierleoni, M., et al., in preparation (2007)
\bibitem{Pollo_VVDS_xi_tech} Pollo, A., Meneux, A. Guzzo, L., et
al. (VVDS Consortium), AA {\bf 439}, 887 (2005)
\bibitem{Pollo2006_xilum} Pollo, A., Guzzo, L., Le F\`evre, O., et
al. (VVDS Consortium), AA {\bf 451}, 409 (2006)
\bibitem{PS74} Press, W.~H., \& Schechter, P.\, ApJ {\bf 187}, 425
  (1974)
\bibitem{Piero_ARAA} Rosati, P., Borgani, S., \& Norman, C., ARAA {\bf
    40}, 139 (2002)
\bibitem{Sheth_Tormen99}Sheth, R.K., \& Tormen, G., MNRAS {\bf 308},
  119 (1999)
\bibitem{Smith03_HOD} Smith, R.E., et al. MNRAS {\bf 341}, 1311 (2003)
\bibitem{Somerville_SAM} Somerville, R.S., Primack, J.R., MNRAS
{\bf 310}, 1087 (1999)
\bibitem{Wang06} Wang, L., Cheng, L., Kauffmann, G., \& De
Lucia, G., MNRAS {\bf 371}, 537 (2006)
\bibitem{White87_CDM}White, S.D.M., Frenk, C.S., Davis, M. \&
  Efstathiou, G., ApJ {\bf 313}, 505 (1987)
\bibitem{Yee_CNOC}Yee, H.K.C., et al., ApJS, {\bf 129}, 475 (2000)
\bibitem{Zehavi04} Zehavi, I., et al., ApJ {\bf 608}, 16 (2004)
\bibitem{Zehavi05} Zehavi, I., Zheng, Z., Weinberg, D.H. et al., ApJ
  {\bf 630}, 1 (2005)
\bibitem{Zucca06_VVDS_LF_types} Zucca, E., Ilbert, O., Bardelli, S.,
  et al. (VVDS Consortium), AA {\bf 455}, 87 (2006)

\end{thebibliography}
}
\end{document}